# Coherence limits in lattice atom interferometry at the one-minute scale

**Authors:** Cristian D. Panda*[1], Matthew Tao[1], James Egelhoff[1], Miguel Ceja[1], Victoria Xu[1†], Holger Müller*[1]

[1]Department of Physics, University of California, Berkeley, 94720, CA, USA.

*Corresponding authors. Email: cpanda@berkeley.edu, hm@berkeley.edu

**Abstract:** In quantum metrology and quantum simulation, a coherent non-classical state must be manipulated before unwanted interactions with the environment lead to decoherence. In atom interferometry, the non-classical state is a spatial superposition, where each atom coexists in multiple locations as a collection of phase-coherent partial wavepackets. These states enable precise measurements in fundamental physics and inertial sensing. However, atom interferometers usually use atomic fountains, where the available interrogation time is limited to around 3 seconds for a 10 m fountain. Here, we realise an atom interferometer with a spatial superposition state that is maintained for as long as 70 seconds. We analyse the theoretical and experimental limits to coherence arising from collective dephasing of the atomic ensemble. This reveals that the decoherence rate slows down markedly at hold times that exceed tens of seconds. These gains in coherence may enable gravimetry measurements, searches for fifth forces or fundamental probes into the non-classical nature of gravity.

---

[†] Present address: Massachusetts Institute of Technology, Cambridge, MA 02139, USA

Non-classical states whose coherence can be maintained for tens of seconds or even minutes are at the forefront of quantum science. For example, optical lattice clocks with atoms trapped for 15 s have enabled 2-3 orders of magnitude gains[1,2] in precision relative to atomic fountain clocks. In atom interferometry, achieving minute-scale coherence times would be instrumental for portable and vibration insensitive gravimetry[3–5] and many applications that are currently out of reach, such as testing the quantum nature of gravity[6–9], or increasingly precise searches of dark energy candidates[10,11]. However, atomic fountains have been limited to 2.3 s by the free-fall times available in 10-meter fountains[12]. Since the height required scales quadratically with time, experiments with fountains measuring hundreds of meters[13–16], sounding rockets[17,18], zero-gravity planes[19], drop towers[20], and in microgravity on the International Space Station[21,22] aim for several seconds of interrogation time, but none are currently proposed to achieve minute-scale coherence.

Initial demonstrations with atom interferometers suspended in an optical lattice[23–26] have observed coherent quantum spatial superposition states for a few seconds, paving the way for the exploration of much longer temporal regimes. Recently, we used the mode of a Fabry-Perot cavity to strongly reduce optical lattice imperfections, extending measurement time to 20 s[27]. However, the exact fundamental or technical limits remained unexplored, limiting achieved coherence.

Here, we elucidate these limits, observing coherences for as long as 70 seconds, the longest-lasting spatial superposition of any massive particles, 30 times longer than observed with atoms in free-fall[12]. Compared to atomic fountains, the mm-scale trajectories in our experiment facilitate maintaining homogeneity of electric, magnetic, and gravitational fields at levels required for precision measurement[28]. We experimentally investigate the sources of decoherence, identify key influences, and develop a framework to describe decoherence in the optical lattice. We study coherence at hold times beyond 20 s and observe significant slowing down in the decoherence rates. Our modelling describes decoherence of spatial superposition states in an optical lattice, unlike previous studies of decoherence of internal superposition states[29].

**A coherent spatial superposition measured after 70 seconds**
The experiment sequence starts by preparing a sample of cold Cesium (Cs-133) atoms (**Figure 1a**) in the magnetically insensitive $m_F = 0$ state of the ground hyperfine manifold of the Cesium atom (see Methods). The atoms are launched upwards by a moving optical lattice (**Figure 1b**) and their axial temperature is further reduced by a Fourier-limited Raman $\pi$-pulse to 80 nK (transverse temperature $T_\perp^0 = 300$ nK). In free fall, counterpropagating laser pulses at 852 nm wavelength stimulate two-photon Raman transitions between the $F = 3$ and $F = 4$ hyperfine ground states [30]. These pulses' intensities and durations are tuned to transfer atoms with a 50% probability ($\pi/2$ pulses). This leads to a coherent beam splitter that separates the atomic matter-wave into two partial wavepackets that move with a differential velocity of $2v_r = \hbar\, k_{\text{eff}}/m_{\text{Cs}} = 7$ mm/s given by the momentum of the 852 nm photons, where $k_{\text{eff}}$ is the laser field wavevector and $m_{\text{Cs}}$ is the Cs atom mass.

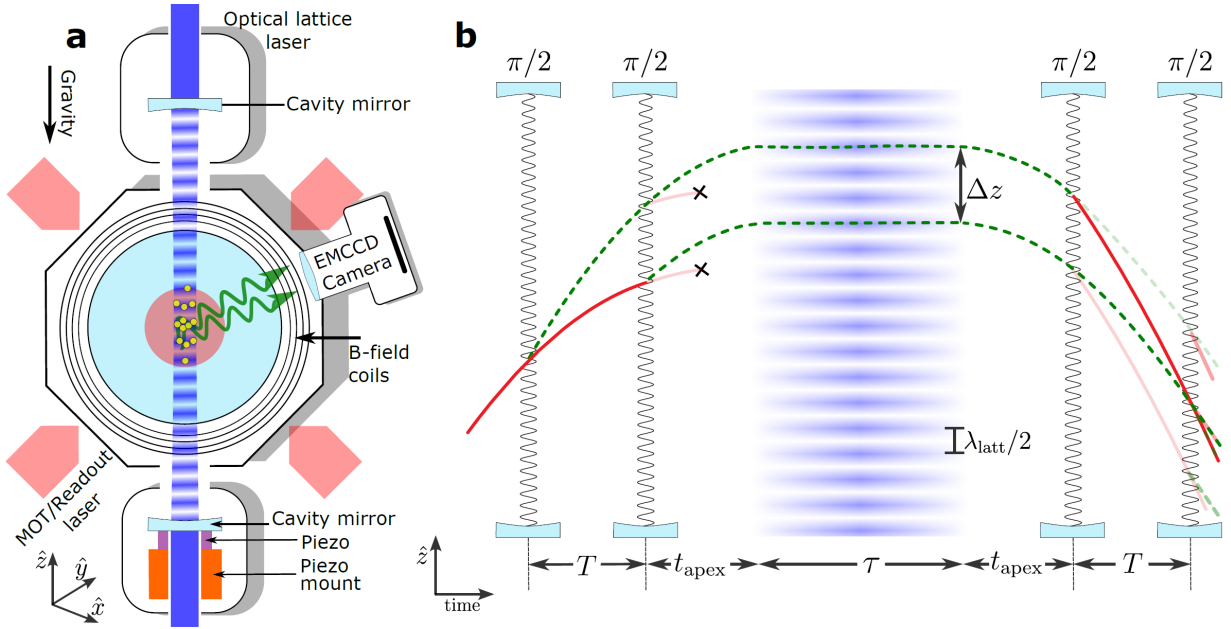

**Figure 1. a, Schematic of the apparatus.** Cs atoms (small yellow disks) are magneto-optically trapped (MOT) at the center of an ultra-high vacuum chamber by 6 laser beams (red, wavelength 852 nm, near resonant with the Cs D2-line). They are loaded into the high-intensity regions of the standing wave of an optical lattice laser (dark blue, wavelength 943 nm), formed by the fundamental mode of a vertically oriented Fabry-Perot cavity (mirrors in light blue). A ring piezo (purple) on a mount (orange) keeps the cavity length constant. Fluorescence photons (green wiggly lines) are detected by an EMCCD camera. **b,** Atom interferometer trajectories. Pairs of $\pi/2$ pulses (wavy vertical lines) separated by time $T$ split, redirect and interfere the launched atomic wavepacket. At the apex of their free-fall trajectory (time $t_{apex}$ after the second beamsplitter), the atomic wavepackets, vertically separated by $\Delta z$, are loaded in the far-detuned optical lattice (horizontal blue stripes, lattice spacing $\lambda_{latt}/2$) where they remain for time $\tau$. The internal atomic state of each interferometer arm is either $F = 3$ (red, solid lines) or $F = 4$ (green, dashed lines) hyperfine levels. Lighter shade lines represent interferometer arms that do not interfere and the two black crosses show arms that are removed using a push laser beam.

Applying a pair of $\pi/2$ pulses, separated by a time $T$, splits the matter-waves four-fold. We select two wavepackets that are moving at a constant separation $\Delta z = 2v_r T$, while sharing the same internal quantum state and external momentum. When they reach the apex, they are adiabatically loaded into the high-intensity regions of a far-detuned optical lattice with a spatial periodicity of $\lambda_{latt}/2$, where $\lambda_{latt} = 943$ nm is the wavelength of the lattice laser, and a trap depth $U_0$. Holding the atoms against Earth's gravitational field requires $U_0 > 4\,E_r$ (experimentally measured, losses are tunneling dominated) for trap lifetimes exceeding 10 s. We use the Cs atom recoil energy at 852 nm, $E_r = m_{Cs}v_r^2/2 = \hbar \cdot 2\pi \cdot 2.0663$ kHz as a unit of energy, even though the lattice wavelength is 943 nm. Operation in microgravity would allow a strong reduction in the trap depth (tunneling between lattice sites can be suppressed with a magnetic field gradient or by using atom-atom interactions to break their degeneracy). After a hold time $\tau$, the atomic wavepackets are adiabatically unloaded and recombined using a final pair of $\pi/2$ pulses.

Due to the resulting interference, the phase difference $\Delta\phi = \phi_t - \phi_b$ accumulated between the top and bottom wavepackets determines the probabilities $P_{3,4} = [1 \pm C \cos(\Delta\phi)]/2$ of detecting an atom in the output ports corresponding to $F = 3$ and $F = 4$, where $C$ is the fringe contrast. The maximum contrast is $C = 0.5$ because only two of the four interferometer outputs interfere. The atom numbers ($N_{3,4}$) in each port, which are proportional to $P_{3,4}$, are measured through fluorescence imaging. We extract the interferometer phase, $\Delta\phi$, from the measured population asymmetry, $A = \frac{N_3 - N_4}{N_3 + N_4} = C \cos(\Delta\phi)$. The primary contribution to interferometer phase is the propagation phase $\Delta\phi_{\text{grav}}^{\text{prop}}$ (see Methods and Extended Data Figures 1 and 2 for other contributions), which accumulates due to the gravitational potential difference $\Delta U_{\text{grav}} = m_{\text{Cs}} g \Delta z$ between the top and bottom interferometer arms during the lattice hold and is obtained by integrating the Lagrangian $\mathcal{L}$ over the classical trajectory[31],

$$\Delta\phi_{\text{grav}}^{\text{prop}} = \frac{1}{\hbar} \left( \int_{t=0}^{\tau} \mathcal{L}^t \, dt - \int_{t=0}^{\tau} \mathcal{L}^b \, dt \right) = \frac{1}{\hbar} \int_{t=0}^{\tau} \Delta U \, dt = \frac{m_{\text{Cs}} g \Delta z}{\hbar} \tau. \quad (1)$$

Over the range of wavepacket separations, hold times, and trap depths that maximize sensitivity to the measurement of accelerations, we find that interferometer noise is consistent with the standard quantum limit (see "Sensitivity and noise" section in Methods and Extended Data Figure 3). The apparatus contains several upgrades over our previous one[27]. Improved sample preparation and a more efficient moving-lattice launch cumulatively increased atom number 40-fold. Lower atomic sample temperature provided 1.4-fold lower contrast loss. A further detuned optical lattice laser (943 nm instead of 866 nm) reduced decoherence from single photon-scattering. Improved stabilization of the intracavity lattice laser intensity increased the atomic lifetime to 14 seconds from 7 seconds. Imaging efficiency was improved through reduction of stray light using blackened radiation shields and the use of an electron-multiplied charge-coupled devices (EMCCD) camera with high (~40% at 852 nm) quantum efficiency and low noise. Laser phase noise has been reduced by a more stable radio frequency (rf) reference. As a result, the interferometer precision to measuring acceleration is $5 \, \mu\text{m/s}^2/\sqrt{\text{Hz}}$, an order of magnitude improved compared to the precision measured in our previous setup after publication of the manuscript[27]. This improved sensitivity is compatible with applications where long measurement times are advantageous, such as high-resolution measurements of gravity from small source masses[10,11] or searches for quantum gravity.[6,7]

In addition, we now observe up to 70 seconds of coherent phase accumulation across the atomic spatial superposition state (**Figure 2**). This interferometer used a pulse separation $T = 0.267$ ms ($\Delta z = 1.9 \, \mu\text{m}$) and peak trap depth of $U_0 = 7 \, E_r$, sufficient to detect several hundred atoms per 70 s experiment cycle. Our fitting routine finds that the contrast of the fringes at 30, 60 and 70 seconds, respectively, is non-zero at the levels of 3.0, 2.6 and 2.9 fit sigma, corresponding to a confidence of 99.1%, 97% and 99.0% respectively (99.9% combined confidence for the 60 and 70 s data) that the atoms exhibit coherence (see Methods, Extended Data Figure 4 and Extended Data Table 1).

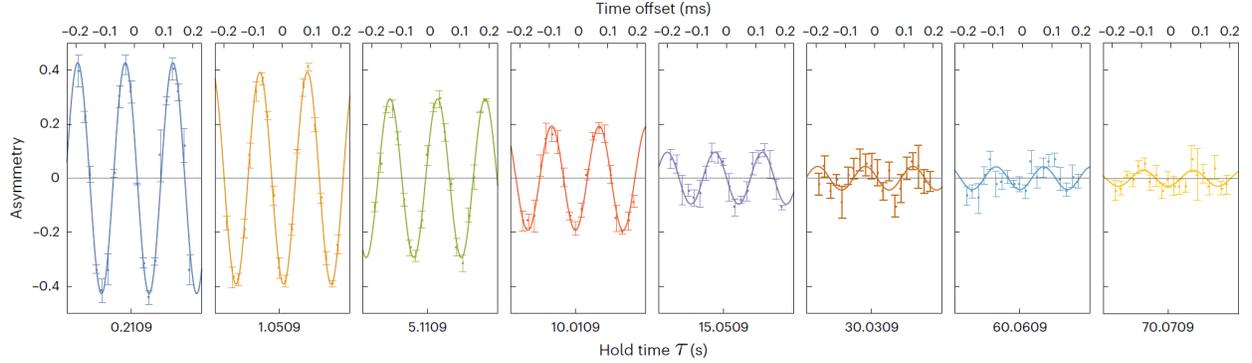

**Figure 2. Interference fringes with over one minute coherence.** Oscillations in the asymmetry are shown as the interferometer phase varies proportionally to the hold time τ. The upper horizontal axis shows offsets from the mean time for each fringe, which is shown on the lower horizontal axis. Each datapoint corresponds to the average of a few experiment cycles. The error bars correspond to 1σ standard error. Solid lines show sinusoidal fits to the data. Fringe offsets are removed.

**Measured lattice decoherence**

We quantify interferometer coherence by extracting the fringe contrast, $C$, from a least-squares fit of the interferometer fringes with contrast, phase, and fringe offset as free parameters. For hold times $\tau < 20$ s, we observe the contrast to decay exponentially with the hold-time, $\tau$: $C = C_0 \text{Exp}(-\tau/\tau_C)$ (**Figure 2a**), where $C_0$ is the interferometer contrast with no hold time ($C_0 \sim 0.5$ typically). When $\tau > 20$ s, we observe a marked slowdown in the contrast decay, which is consistent with our model and described in more detail below and in Methods. Because of the large error bars in the data, the exact decay rate is hard to determine but is compatible with a simple exponential decay, $C_1 \text{Exp}(-\tau/\tau_{C1})$, with a time constant $\tau_{C1} \gtrsim 58$ s (1 sigma confidence in a fit where the datapoints are weighted by their error bars). We will see later that this slowdown results from the fact that the fastest-moving atoms of the thermal ensemble leave the trap during the first 20 s.

We initially focus our analysis on the contrast decay term, $\tau_C$, that dominates at $\tau < 20$ s. The decay time constant $\tau_C$ is measured to be inversely proportional to the atomic wavepacket separation, $\Delta z$, and the trap depth, $U_0$ (Figure 3b).

$$1/\tau_C = U_0 \Delta z / \kappa_0. \quad (2)$$

Here, $\kappa_0 = 110 \ \mu\text{m} \cdot \text{s} \cdot E_\text{r}$ was the global contrast decay parameter we measured initially (Figures 3b-d). As described above, we have since optimized atom sample colling, which improved $\kappa$ by a factor of 1.4 to 160 $\mu\text{m} \cdot \text{s} \cdot E_\text{r}$, which is used to record fringes with residual coherence after up to 70 s, as shown in Figure 2.

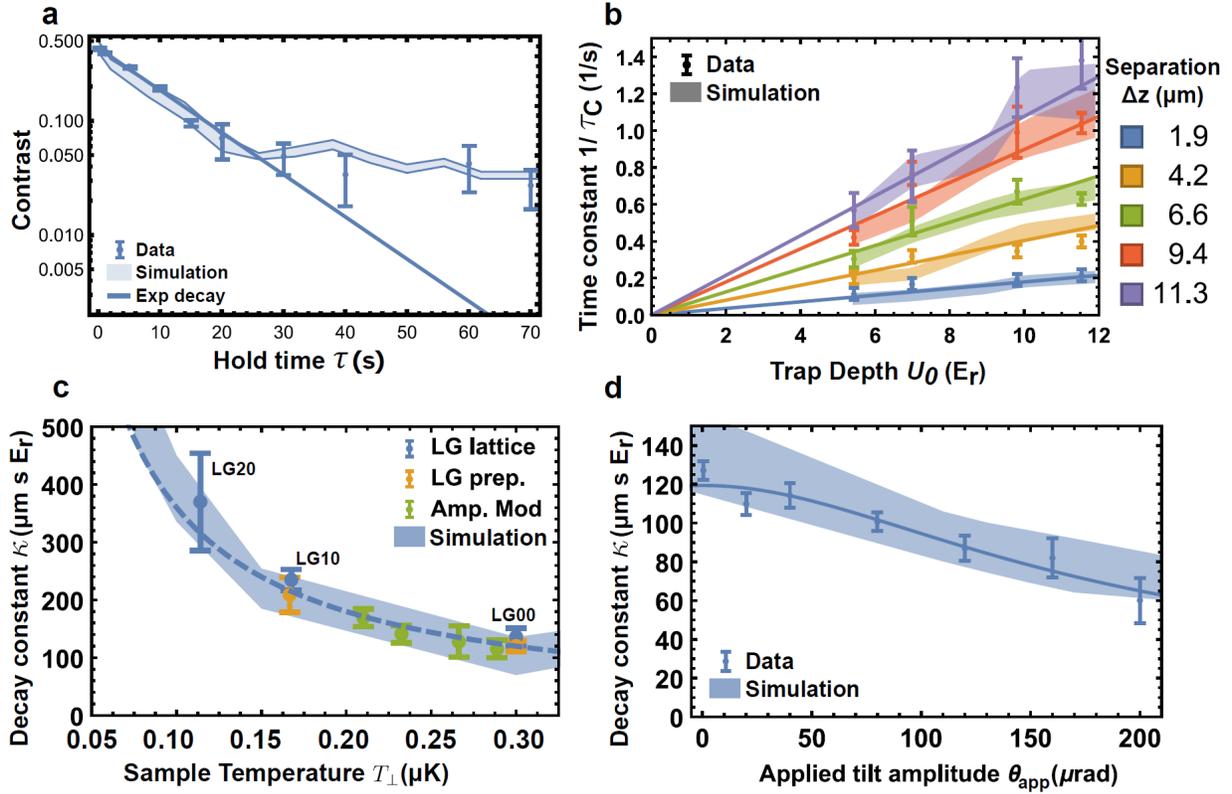

**Figure 3. Decoherence during optical lattice hold. a,** Contrast decays exponentially vs hold time $\tau$ with time constant $\tau_C$, $C = C_0 \text{Exp}[-\tau/\tau_C]$, for $\tau < 20$ s (least squares fit shown with solid line) and at a slower rate for $\tau > 20$ s. **b,** $\tau_C$ is inversely proportional to wavepacket separation, $\Delta z$, and trap depth, $U_0$ (measured in recoil energy units $E_r$): $1/\tau_C = U_0 \Delta z / \kappa$ (least squares fit to data shown with solid line). Data was recorded with $\tau$ in the range of 0 to 5 seconds. **c,** The global contrast decay parameter $\kappa$ increases when using atomic ensembles that are colder transversally. Experimental data with three different methods of achieving lower transverse atom temperature $T_\perp$ are shown: using high order LG10 and LG20 modes during lattice hold (LG lattice), using high order LG10 and LG20 modes only during state preparation (LG prep) and using amplitude modulation to "boil off" hot atoms (Amp. Mod.). The dashed line shows $\kappa(T_\perp) = \kappa_0 * T_\perp^0 / T_\perp$, where $\kappa_0$ and $T_\perp^0$ are the typical contrast decay parameter and transverse atom temperature. **d,** Decrease of the contrast decay constant $\kappa$ with magnitude of applied tilt $\theta_{\text{app}}$. The line is a fit to equation $\kappa(\theta_{\text{app}}) = \kappa_0 / \sqrt{1 + (\theta_{\text{app}}/\theta_0)^2}$, where $\theta_0$ is the residual environmental noise. All data points are shown with error bars corresponding to $1\sigma$ standard error. The color bands correspond to $1\sigma$ (68%) Gaussian confidence bands representing the statistical uncertainty of a numerical simulation that estimates atomic ensemble dephasing in the presence of residual atomic motion and lattice tilt noise.

To understand the physical mechanisms that limit the coherence time, we tested many modifications to the experiment. We found that $\kappa$ was robust against several strong changes to the setup. This includes using higher-quality cavity mirrors (surface rms roughness <1 A) as opposed to standard dielectric mirrors as well as locating the cavity waist near the atoms, so that the lattice potential is unchanged between the two interferometer arms. We also found that lattice laser imperfections (frequency noise, broadband emission, background scatter, imperfect cavity coupling), the alignment of the lattice with gravity, atom density and axial temperature (which is much lower than the transversal temperature due to axial velocity selection), lattice laser beam properties (polarization, wavelength), the symmetry of interferometry pulses and magnetic fields (see Extended Data Table 2 in Methods for details), did not affect $\kappa$.

However, $\kappa$ is improved up to 3-fold when using higher order Laguerre-Gaussian (LG) modes for the optical lattice hold (**Figure 2c** and Supplementary Movie 1). One possible explanation is that loading into higher-order cavity modes selects a subset of atoms with a lower transverse temperature, $T_\perp$. The radially symmetric LG10 and LG20 modes have a central intensity peak that is narrower than the fundamental LG00 Gaussian mode (widths of 56% and 43% respectively at half maximum). The selection occurs when the narrower lattice beam captures only atoms near the center of the atom cloud. (The low intensity "rings" of the higher-order modes are too weak to trap atoms and have poor overlap with the atom sample, so atoms only get trapped in the central peaks.) We corroborate this interpretation by temperature measurements of the trapped atoms using time-of-flight (TOF) spectroscopy, which show that centrally trapped atoms are the coldest (Fig. 3c). We believe that the reason for the lower temperature is twofold: first, a spatial variation of the intensity of cavity-derived Raman-sideband cooling beams, and second, that the fastest-moving atoms disappear from the trapping region in the 38 ms interval between laser cooling and lattice loading.

To further corroborate that the lower transverse temperature is the cause for improved coherence, we have used the LG10 mode to temperature-select the atomic sample while switching back to the LG00 fundamental mode during interferometry hold. A TOF measurement confirms that the LG10 mode selects an atom sample that is 1.8 times colder. We observe a 1.8-fold increase in the contrast decay constant $\kappa$ (**Figure 2c**).

We further confirm that coherence depends on sample temperature by "boiling off" the most energetic atoms by amplitude-modulating the lattice during the initial 500 ms of the hold. The lattice trap depth was set to $U_0 = 14\,E_\mathrm{r}$, corresponding to axial trap frequency $\omega = 2\pi \cdot 14$ kHz. The modulation has an amplitude of 5% of the trap depth at a frequency that is twice the trap frequency, $2\omega$. This reduces the sample temperature to 70% of the initial value, as measured through TOF. We observe that this reduces the number of remaining atoms 3-fold and leads to an increase in contrast (**Figure 2c**).

In addition to the atom temperature, we observe that oscillatory tilts $\theta(t)$ of the trapping laser beam with respect to the vertical gravitational axis are a strong influence on $\kappa$. To demonstrate this, we shake the experiment (which is floating on minus-K vibration isolators) along the $-\hat{x}$ axis (Fig. 1) using a voice coil, typically at $2\pi \cdot 600$ Hz. The resulting tilt amplitude $\theta_\mathrm{app}$ was quantified by measuring the position of the cavity transmitted laser beam on a quadrant photodetector. Fringes were measured for varying amplitudes of applied tilt (**Figure 2d**). The resulting $\kappa$ was fitted to a model $\kappa(\theta_\mathrm{app}) = \kappa_0 / (1 + (\theta_\mathrm{app}/\theta_0)^2)^{1/2}$, where $\theta_\mathrm{app}$ is the applied tilt and the fitted $\theta_0 = 120\,\mu$rad is the residual environmental tilt.

We independently quantify the environmental tilt spectrum in our experiment $\theta_0$. The spectrum consists of multiple peaks, with the largest components in the few hundred hertz band, with a total rms tilt of $200 \pm 150\,\mu$rad, consistent with that obtained from the fit above.

**Understanding decoherence due to atomic ensemble dephasing**

The above observations are compatible with a model that connects contrast decay with time-dependent tilts and residual (thermal) atomic motion.

In addition to the gravitational phase difference $\Delta\phi_{\text{grav}}^{\text{prop}}$ (Eq. 1), each arm of the interferometer accumulates phase due to the lattice lightshift (AC Stark shift) from the optical potential $U_{\text{latt}}$. $U_{\text{latt}}$ has a Gaussian profile in the transverse $xy$-plane and depends sinusoidally on $z$: $U_{\text{latt}} = U_0 \text{Exp}\left(-\frac{x^2+y^2}{2w_0^2}\right)\text{Sin}^2\left(\frac{2\pi z}{\lambda_{\text{latt}}}\right)$, where $w_0 \approx 760\ \mu m$ is the radius of the cavity mode. For an atom at rest at the center of a lattice with $U_0 = 7\ E_r$, $\phi_{\text{latt}}^{\text{prop}} = 6.4$ Mrad after $\tau = 70$ seconds.

Ideally, the trapping potentials for the top and bottom interferometer arms are identical, so that $\Delta\phi_{\text{latt}}^{\text{prop}} = \phi_{\text{latt}}^t - \phi_{\text{latt}}^b \cong 0$. However, any difference in the optical potential between the top and bottom arms, $\Delta U_{\text{latt}} = U^t - U^b$, will lead to a differential phase shift $\Delta\phi_{\text{latt}}^{\text{prop}} \propto \Delta U_{\text{latt}} \neq 0$. In addition, $\Delta U_{\text{latt}}$ moves the top and bottom partial wave packets relative to each other, so that they do not overlap exactly at the end of the interferometer, which causes a separation phase contribution, $\Delta\phi_{\text{latt}}^{\text{sep}}$ [32]. The lattice phase shift is thus given by $\Delta\phi_{\text{latt}} = \Delta\phi_{\text{latt}}^{\text{prop}} + \Delta\phi_{\text{latt}}^{\text{sep}}$.

Despite taking great care to minimize experimental imperfections, as described above, sources of differential phase shift remain. In particular, time-dependent tilts due to vibrations faster than the trap frequency cause the top and bottom lattice sites to oscillate transversally with a differential amplitude proportional to $\Delta z$: $a_{\text{tilt}}(t) = \theta(t) \cdot \Delta z$. This motion causes a differential potential, $\Delta U_{\text{tilt}}(t) \propto \Delta z \cdot \theta(t) \cdot U_0$. To first order, the equations above give a linear phase shift as a function of tilt amplitude, trap depth, separation and hold time

$$\Delta\phi_{\text{latt}}^{\text{tilt}} \propto \Delta z \cdot \theta(t) \cdot U_0\ \tau. \tag{3}$$

The lattice phase shift also depends on the distribution of atomic trajectories in the ensemble, which motivates the transverse temperature dependence observed above. We focus here on motion in the transverse $xy$-plane, where the atoms follow orbital motion around the center of the optical trap. These trajectories can be treated semi-classically since the size of the atomic wavepackets (~500 nm) is much smaller than the size of the trap potential (waist 760 $\mu m$) and the transverse motion is much slower than the axial motion. A representative atomic-motion timescale is given by the transverse harmonic trap frequency $\omega_\perp = 2\pi \cdot 2.8$ Hz, much slower than the measured cavity tilts which typically occur at frequencies $>2\pi \cdot 100$ Hz. However, the trajectories have appreciable non-harmonic components, because of the long tails of the Gaussian potential. The spread in initial velocities and positions causes the lattice phase to accumulate at different rates for different atoms. This means that a hotter atom cloud has a wider phase dispersion

$$\delta(\Delta\phi_{\text{latt}}) \propto \Delta z \cdot \theta(t) \cdot U_0\ \tau\ T_\perp. \tag{4}$$

We simulate the atomic trajectories numerically to estimate the proportionality factors (see Methods, Extended Data Figures 5, 6 and Extended Data Table 3). The simulations result in a Lorentzian phase distribution across the sample, $P(\Delta\phi) = \frac{1}{\pi} \cdot \frac{\delta(\Delta\phi_{\text{latt}})}{\Delta\phi^2 + [\delta(\Delta\phi_{\text{latt}})]^2}$, with a width $\delta(\Delta\phi_{\text{latt}})$ that scales like Eq. (4). We use a density matrix formalism (see Methods) to show that this Lorentzian phase distribution results in an exponential decay of the ensemble contrast[33] with $\delta(\Delta\phi_{\text{latt}})$

$$C = C_0 \exp[-\delta(\Delta\phi_{\text{latt}})]. \tag{5}$$

Thus, the combination of differential phase shift due to tilts (Eq. ( 3 )), phase dispersion due to residual atomic motion (Eq. ( 4 )) and ensemble collective dephasing (Eq. ( 5 )) predicts a decoherence model

$$C = C_0 \exp[-\Delta z \cdot U_0 \cdot \tau/\kappa], \qquad (6)$$

where $\kappa \propto (\theta(t) \, T_\perp)^{-1}$. The contrast decay model predictions are quantified through numerical simulation, which results in quantitative agreement to the experimental data for timescales $\tau < 20$ s (**Figure 2**).

For timescales $\tau > 20\,s$, we observe slowing down in contrast loss that is consistent between the experiment and simulation (Fig. 3a). After 20 seconds, the simulation shows that most atoms in high energy trajectories near the edge of the trap potential have left the interferometer due to tunneling. Therefore, the remaining atoms are in low energy trajectories near the flat region at the center of the Gaussian trap potential. These atoms are less sensitive to trap motion caused by tilts than high energy atoms and therefore experience reduced decoherence. These dynamics explain why contrast loss slows down for $\tau > 20\,s$. To further check these conclusions, we ran the simulation with tunnelling disabled, which produces a higher rate of contrast loss, as expected, since high-energy atoms remain in the shallow parts of the trap (Extended Data Figure 7).

**Discussion**

The model is useful for estimating dephasing due to a variety of experimental imperfections. In addition to decoherence due to tilts considered above, we have also used our model to estimate that decoherence caused by several other effects is 1-2 orders of magnitude too weak to matter here. Effects considered include atomic motion from the Coriolis force, asymmetric transverse kicks on the atoms from the first and second sets of Raman beamsplitters, curvature of the optical lattice when the atoms are away from the waist, parasitic optical lattices due to residual lattice laser light resonant with the cavity and imperfections in the optical potential due to speckle.

Our models and experimental tests suggest further improvement in atom numbers and readout noise, reduction of atom sample temperature and reduction of oscillatory tilts may lead to spatial superposition states with ultra-long coherence, exceeding minutes. These increases in coherence are landmark results towards applications such as gravimetry[3–5], searches for fifth forces[10,11], or fundamental probes into the non-classical nature of gravity[6,7,34].

**Acknowledgments:** We thank J. Axelrod and A. Reynoso, for experimental assistance; J. Lopez for technical support; L. Clark, D. Carney, N. Gaaloul, M. Jaffe, P. Haslinger, Z. Pagel, G. Premawardhana, J. Roth, A. Singh, J. Taylor for valuable discussions.

This material is based on work supported by the

> National Science Foundation grants 1708160 and 2208029 (HM)
>
> Department of Defense Office of Naval Research grant N00014-20-1-2656 (HM)
>
> Jet Propulsion Laboratory (JPL) grants 1659506 and 1669913 (HM)

**Author Contributions Statement:** CDP, JE, MT and MC built the apparatus and collected measurements. CDP analyzed data. VX contributed to building the apparatus and initial investigations into the source of decoherence. CDP and HM wrote the original draft. HM conceptualized and supervised the experiment. All authors contributed to the review and editing of the manuscript.

**Competing Interests Statement:** The authors declare no competing interests.

## Methods

Atom sample preparation and detection

To describe the experiment in detail, we define a coordinate system in which $\hat{z}$ points upwards and is aligned to Earth's gravity. The optical cavity axis is approximately aligned with $\hat{z}$. We define $\hat{x}$ to lie in the horizontal plane, aligned with the direction of the MOT magnetic field coils, such that $\hat{y} = \hat{z} \times \hat{x}$ is approximately along the direction of the camera (Fig. 1a).

Initial trapping of the atoms in a 2-dimensional MOT (2D MOT) takes place in a glass vacuum chamber (not shown in Fig. 1a) which contains Cs vapor at room temperature. Four laser beams, 12.8 MHz red-detuned from the F=4 to F'=5 line in the Cs D2 manifold and the magnetic field generated by four coils create a 2D MOT. The 2D MOT axis is aligned with a differential pumping aperture tube that connects to the primary experimental chamber (octagon in Fig. 1a). The primary chamber is evacuated with ion pumps and titanium sublimation pumps to a pressure of $6 \times 10^{-11}$ torr. The Cs atoms are loaded (loading time 700 ms) into a 3D MOT formed by 6 independent laser beams (wavelength 852 nm, 12.8 MHz red detuned from the F=4 to F'=5 line) with a beam diameter of ~3 cm and a magnetic quadrupole field. Three additional Helmholtz coils are used to zero the environmental magnetic field along the $\hat{x}$, $\hat{y}$ and $\hat{z}$ directions. The MOT atoms load at the spatial location where net environmental plus applied magnetic field is 0.

After the 3D MOT stage, the magnetic quadrupole field is turned off and the atoms are laser cooled by optical molasses to 10 $\mu$K. The atoms are then Raman Sideband Cooled (RSC) in a 3D optical lattice (trap depth ~50 $E_r$, wavelength 852 nm, ~50 GHz red-detuned from D2), where two of the laser beams are derived from the TEM00 cavity mode and the other two are provided through the side window. The resulting RSC temperature is 0.3 $\mu$K, near the Cs recoil limit.

The atoms are then transferred into a far off resonant 1D lattice formed by the TEM00 mode of the optical cavity (trap depth ~70 $E_r$, wavelength 943 nm). RSC leaves atoms in the $F = 3, m_F = 3$ stretched state[35]. The atoms are then transferred using microwave rapid adiabatic passage (9 ms ramp) followed by a microwave $\pi$-pulse to the magnetically insensitive $F = 3, m_F = 0$ state.

To launch the atoms vertically upwards, the atoms are transferred into a vertically oriented lattice formed by two laser beams that enter the experiment diagonally, along the direction of the MOT laser beams. The frequency of one of the laser beams is swept to launch ~80% of the atoms upwards into free fall. The launch velocity is typically set to 196 mm/s, such that the atoms reach the apex of their trajectory after 20 ms of free-fall time, corresponding to a launch height of 1.96 mm. A Doppler sensitive Raman $\pi$-pulse along $\hat{z}$ (Gaussian time profile with $1\sigma$=20 $\mu$s) restricts the longitudinal atomic velocity to $v_\parallel = 0.6\, v_r$ (Gaussian $1\sigma$), corresponding to a temperature of 80 nK and $2\pi \cdot 8$ kHz (Gaussian $1\sigma$) Doppler shift.

To obtain a pure sample, atoms that did not take part in the population transfer need to be removed. This is accomplished by a laser beam resonant with the D2 line (either $F = 4$ to $F' = 5$ or $F = 3$ to $F' = 2$) which travels approximately along the $-\hat{x} + \hat{z}$ direction and is used to push atoms with internal state that is either $F = 4$ or $F = 3$, respectively. It is also used at the beginning of the interferometer to push away two of the interferometer arms that are transferred in the unwanted momentum and hyperfine state (Fig. 1b).

For detection after the interferometer sequence, this same laser is used to push the atoms in $F = 4$ over a distance of ~1 mm relative to the $F = 3$ atoms, so that the two populations can be separately detected by fluorescence within one camera image. Fluorescence is excited using the MOT laser beams, blocking the ones along the $\hat{x}$ direction to reduce stray light. The detection light is typically 5.6 MHz red-detuned from the $F = 4$ to $F' = 5$ line and the atoms are imaged for 4-10 ms before imperfections in the laser light intensity profile or power imbalance push away the atom sample. This scheme measures both interferometer outputs simultaneously on the same image, minimizing detection noise due to fluctuations in the intensity or detuning of the detection light or variations in background stay light.

The camera is placed at a 45º angle with respect to the horizontal, along the $\hat{z} + \hat{x}$ direction and is pointed towards the atomic cloud. A high-NA objective (f/0.95 aperture, Navitar MVL50HS) is used to collect and image the fluorescence on the surface of the EMCCD sensor. The EMCCD gain is typically set to 0, except for long hold time measurements ($\tau > 20$ s), where the EMCCD gain is set to values up to 300x to image lower numbers of atoms ($N < 1000$).

Optical cavity and lattice imperfections

The optical Fabry-Perot cavity is formed by two curved mirrors (20 m radii of curvature) that are separated by $L = 37.5$ cm. This results in a cavity waist size of $w_0 = 760$ $\mu$m and a Rayleigh length of $z_R = 1.92$ m at the lattice wavelength $\lambda_{latt}$ =943 nm. The cavity waist is located near the center of the atom cloud, which ensures that the diameter of the lattice at the top and bottom interferometer arms is as similar as possible. This minimizes dephasing due to variations in the transverse diameter of the trap potential between the top and bottom interferometer arms. Numerical simulations suggest that contrast decay due to the remaining small potential difference is now at least three orders of magnitude lower than in the previous experiment configuration, when the atoms were located 19 cm away from the waist. Therefore, after moving the atoms near the cavity waist, the level of contrast decay due to cavity beam divergence is negligible at current sensitivity.

The mirror substrates are 1 inch diameter and superpolished to a root-mean square (rms) surface roughness of <1 Å. They are dielectric coated for 97.0 % reflectivity at $\lambda_{Raman}$ =852 nm, 98.3 % at $\lambda_{latt}$ =943 nm and 93.5 % at $\lambda_{tracer}$ =780 nm. The linewidth of the cavity is 2.2 MHz at 943 nm, providing ~59 times power enhancement of the coupled light. The transverse mode spacing is 20 MHz, sufficiently large compared to the cavity linewidth so that most of the incoming light is coupled into a single transverse cavity optical mode.

Laser frequency stabilization scheme

A reference laser is frequency-stabilized to the Cs D2 line in a vapor cell[36]. All the other frequency stabilization schemes described below use Pound-Drever-Hall (PDH), where the sidebands are provided by an electro-optical modulator (EOM - typical modulation frequency ~20 MHz). The length of a transfer cavity is stabilized with a piezo to the reference laser. The Raman laser ($\lambda_{Raman} = 852$ nm) and tracer laser ($\lambda_{tracer} = 780$ nm) are referenced to this transfer cavity. Both lasers enter the experiment science cavity from the bottom (along $+\hat{z}$), where a set of dichroic mirrors allow their reflection and transmission to be separately monitored by photodetectors. The length of the experiment science cavity described above is stabilized to the tracer laser against vibrations and slow drift due to mechanical expansion of the vacuum chamber.

Light from the lattice laser ($\lambda_{latt} = 943$ nm) is split into two paths. A small fraction (100 $\mu$W) enters the "science" cavity through the top mirror (along $-\hat{z}$) and is used to stabilize the lattice laser to the length of the "science" cavity. The remaining light is amplified by a tapered amplifier and enters the cavity from the bottom mirror (along $+\hat{z}$). The intra-cavity power is monitored by a photodetector placed at the transmission port and is used to stabilize the intra-cavity intensity by adjusting the RF voltage amplitude driving an acousto-optic modulator (AOM). The trap depths quoted throughout the paper are calculated by scaling the measured power of this photodetector.

Lattice properties and loading

The lattice laser used in these experiments is $\lambda_{\text{latt}} = 943$ nm, 49 nm red-detuned from the D1 transition and 91 nm red-detuned from the D2 transition. Compared to our previous work, where $\lambda_{\text{latt}} = 866$ nm with single photon scattering lifetime of $2\pi/\Gamma_{sp} = 2\pi \cdot 11$ s, this larger detuning reduces single photon decoherence to $2\pi/\Gamma_{sp} = 2\pi \cdot 120$ s for $U_0 = 5\, E_{\text{r}}$. This level is negligible for all measurements described here.

Near the apex of the atoms' free-fall trajectories, the optical lattice is turned on or off using a smooth exponential curve with time constant of 400 μs. This is fast compared to the transverse motion of the atoms (characteristic harmonic frequency $\omega_\perp = 2\pi \cdot 2.4$ Hz for $U_0 = 5\, E_{\text{r}}$), but slow compared to the axial motion (characteristic harmonic frequency $\omega_\parallel = 2\pi \cdot 8.2$ kHz for $U_0 = 5\, E_{\text{r}}$). This means that the atoms adiabatically load axially into the high-intensity regions of the optical lattice, which enforces quantization of the atomic wavepacket positions during the hold.

To probe this structure, we measure interferometer fringes when finely changing the initial interferometer arm separation by varying $T$: $\Delta z = v_r T$. We then perform a Fourier decomposition of the measured fringes to determine the primary frequency components. We observe frequency components from gravitational potential differences that necessarily correspond to integer multiples of lattice sites: $\Delta z = n \cdot \lambda_{\text{latt}}/2$, where $n$ is an integer number (Extended Data Figure 1).

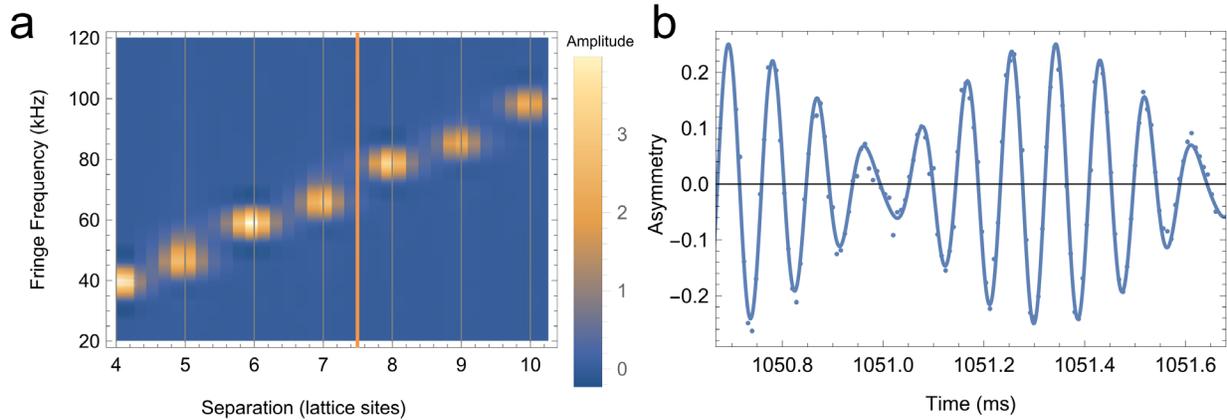

**Extended Data Figure 1. Atoms load into integer multiples of the lattice period. a,** Fourier spectral decomposition components of the measured fringes as a function of separation. Yellow represents a higher power density at the specific fringe frequency. **b,** fringe data when the initial wavepacket separation equals 7.5 lattice sites and fit (solid line) to a sum of two sine functions with frequencies corresponding to separations of 7 and 8 lattice sites. This data suggests that the lattice enacts a second beam splitter, such that the partial atomic wavepacket loads evenly into adjacent lattice sites with 7 and 8 lattice site separations.

Interferometer phase and contrast

The interferometer phase is the sum of two terms: $\Delta\phi = \Delta\phi_L + \Delta\phi_{\text{prop}}$, where $\Delta\phi_L$ results from the interaction of atomic wavepacket with the Raman laser[37] and $\Delta\phi_{\text{prop}}$ is the propagation phase. Each of the four applied laser pulses adds a term $\phi_i$ to this phase that is proportional to the atom's position. In our interferometer,

$$\Delta\phi_L = \phi_L^t - \phi_L^b = (\phi_1 - \phi_2) - (\phi_3 - \phi_4) = k_{\text{eff}} g\, T(T + 2t_{\text{apex}}), \qquad (1)$$

where $g$ is the local gravitational acceleration and indexes $\phi_L^t, \phi_L^b$ denote the phase corresponding to the top and bottom arms, respectively. The total interferometer phase due to Earth's gravitational acceleration $g$ amounts to

$$\Delta\phi = \Delta\phi_{\text{prop}} + \Delta\phi_L = \frac{m_{\text{Cs}} g \Delta z}{\hbar}\tau + k_{\text{eff}} g\, T(T + 2t_{\text{apex}}). \qquad (2)$$

To scan interferometer fringes, we can vary either (a) the hold time, $\tau$, or (b) the rate at which the Raman laser frequency is ramped so that it remains on-resonance with the atoms during free-fall, $\alpha = g * k_{\text{eff}} = 2\pi \cdot 23$ kHz/ms for the D2-line of cesium. Varying $\tau$ results in a fringe frequency that is given by

$$\omega_{\text{prop}} = \frac{\partial(\Delta\phi)}{\partial\tau} = m_{\text{Cs}} g \Delta z / \hbar. \qquad (3)$$

As shown in Extended Data Figure 2a, the adiabatic loading in the optical lattice forces the wavepacket separation to be an integer number of lattice sites, $\Delta z = n \cdot \frac{\lambda_{\text{latt}}}{2}$, where $n$ is an integer. When $\alpha$ is varied, the fringe frequency is

$$\omega_L = k_{\text{eff}} \frac{\partial(\Delta\phi)}{\partial\alpha} = k_{\text{eff}} T(T + 2t_{\text{apex}}). \qquad (4)$$

There is no quantization of scanned parameters in this case (Extended Data Figure 2b).
Additional phase terms can be present due to imperfections in the optical lattice, such as beam curvature, scatter, or environmental magnetic and electric fields. Careful characterization of such systematic contributions is important in the context of future high-precision measurements.

The change in fringe contrast as a function of separation is shown in Extended Data Figure 2c for a hold time of $\tau \approx 1\,s$ and is independent of whether $\alpha$ or $\tau$ are scanned. In addition to the slowly decaying exponential envelope discussed in detail in the main text, we also observe variation in the contrast at shorter separation length scales. The contrast varies as a function of $T$ with a fast sinusoidal component given by lattice spacing, $\lambda_{\text{latt}}/(2v_r)$. In addition, the contrast decay curve exhibits a smaller amplitude (4% typically) sinusoidal component with periodicity given by the least common multiple of the lattice laser and tracer laser wavelengths, $\lambda_{\text{latt}}$ and $\lambda_{\text{tracer}}$. This variation is due to dephasing from the interference of the lattice and tracer lasers. The separations used in Fig. 3 are chosen to minimize these additional dephasing mechanisms and only probe the slow exponential decay of contrast described in the main text. We note that this data was taken before replacing the lattice laser, so $\lambda_{\text{latt}}$=866 nm in Extended Data Figure 2 only.

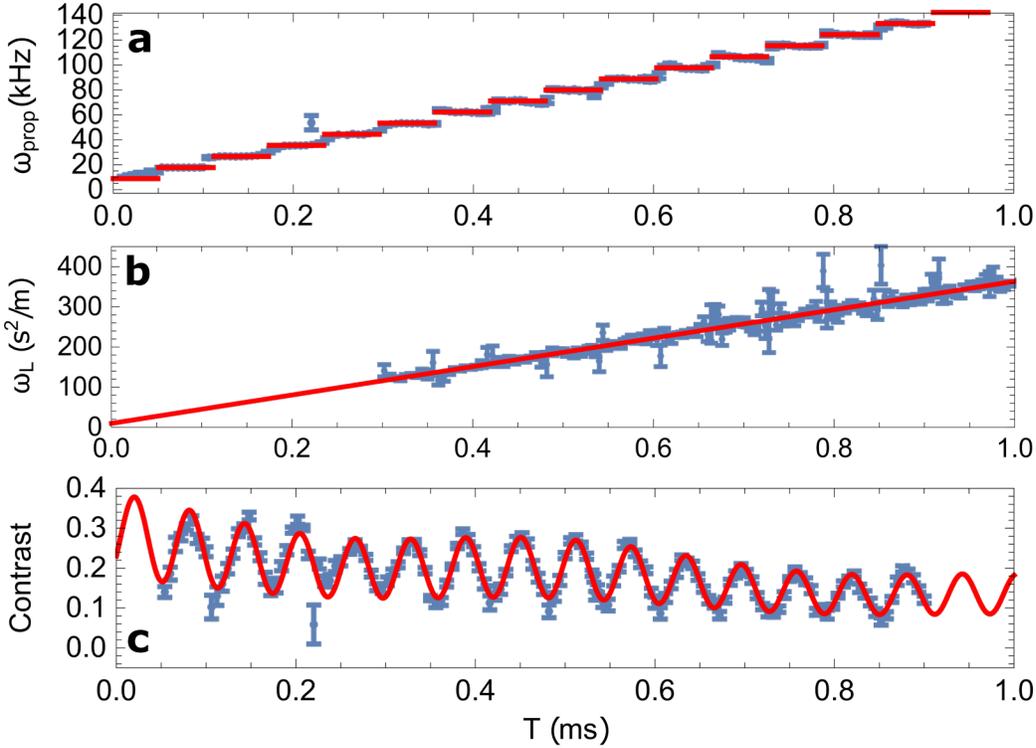

**Extended Data Figure 2. Measured interferometer fringe vs scan parameter. a,** fringe frequency from propagation, $\omega_{\text{prop}}$, obtained by scanning the hold time $\tau$. **b,** fringe frequency from laser interactions, $\omega_L$, obtained by scanning the pulse separation $T$. Error bars correspond to $1\sigma$ (68%) Gaussian confidence intervals. Red lines correspond to the expected analytical fringe frequency as shown by Eqs. 3 and 4. **c,** Contrast loss as a function of separation time, $T$. The fitted interferometer contrast varies as a function of $T$ with a fast sinusoidal component given by lattice spacing times the recoil velocity, $\lambda_{\text{latt}}/2 \cdot v_{\text{rec}}$. In addition, the contrast decay curve exhibits a smaller amplitude (4% typically) sinusoidal component with periodicity given by the least common multiple of the lattice laser and tracer laser wavelengths, $\lambda_{\text{latt}}$ and $\lambda_{\text{tracer}}$, which is due to dephasing from the interference of the lattice and tracer lasers. The red line fits the data using free parameters for the amplitudes of the exponential decay, amplitude of the higher frequency sinusoidal function and the amplitude of the lower frequency variation.

Sensitivity and noise

We compute the standard quantum limit (SQL) of the gravitational sensitivity from

$$\delta g_{\text{SQL}} = \sqrt{2}\hbar/(m_{\text{Cs}} \cdot C \cdot \Delta z \cdot \sqrt{\tau} \sqrt{N \cdot \tau/(\tau + T_{\text{dead}})}), \quad (5)$$

where $T_{\text{dead}} = 1.8\,s$ typically is the time between experimental cycles and the contrast $C$ is measured from the fitted fringes.

The detected atom number $N$ is extracted from the EMCCD images signal $S$ as $N = S/(\Gamma_{\text{eff}} \cdot T_{\text{exp}} \cdot (\Omega/(4\pi)) \cdot \epsilon \cdot \mathcal{A})$, where $\Gamma_{\text{eff}}$ is the photon scattering rate, $T_{\text{exp}}$ is the imaging exposure time, $\Omega$ is the solid angle, $\epsilon$ is the camera signal-to-photoelectron conversion ratio per image area that the signal is summed over, $\mathcal{A}$. $T_{\text{exp}}$ is set by the camera software and $\mathcal{A}$ is calculated from the number of pixels and pixel size. The effective image pixel size is measured by dropping atoms and fitting the resulting atom trajectories to a quadratic function with the known gravitational acceleration $g$. We have found that the fitted fringe phase and contrast are independent of the image integration area $\mathcal{A}$, which is therefore kept fixed for all images within a fringe.

The scattering rate is $\Gamma_{\text{eff}}(s) = s/(1 + s + (2\Delta/\Gamma)^2 ) \cdot \Gamma/2$, where $s = I/I_{\text{sat}}$ is the saturation parameter, $\Delta$ is the laser detuning from the the cesium D2 transition and $\Gamma = 2\pi \cdot 5.2$ MHz is the D2 transition linewidth. We vary the detection laser intensity and fit to the previous equation to find $\Gamma_{\text{eff}} = (0.65 \pm 0.1) \cdot \Gamma/2$ when using typical detection laser intensity. The solid angle is $(\Omega/(4\pi)) = 0.3 \pm 0.1\%$, estimated from the aperture of the collection lens of 2.4 cm, situated at 11.5 cm from the atoms. The camera efficiency $\epsilon = 0.073 \pm 0.02/\mu\text{m}^2$ is measured by shining a beam of same wavelength (852 nm) with known power (measured with a calibrated photodetector) onto the camera. All-in-all, this procedure allows the measurement of the atom number $N$ with an uncertainty of about 50%.

To check if the measured sensitivity is consistent with that expected from atom statistics at the SQL, we acquire fringes with different $\tau, U, \Delta z$. Using the procedure described above, we obtain the atom number $N$ as a function of $\tau$ for fringes with $U = 12\ E_{\text{r}}$ (Extended Data Figure 3a). We find agreement between the measured and estimated SQL sensitivity values (shown in Extended Data Figure 3b for $U = 12\ E_{\text{r}}$ and $\Delta z = 6.6\ \mu\text{m}$). The best sensitivity when measuring $g$ is around $5\ \mu\text{m/s}^2/\sqrt{\text{Hz}}$ and can be achieved for a relatively large range of experimental parameters.

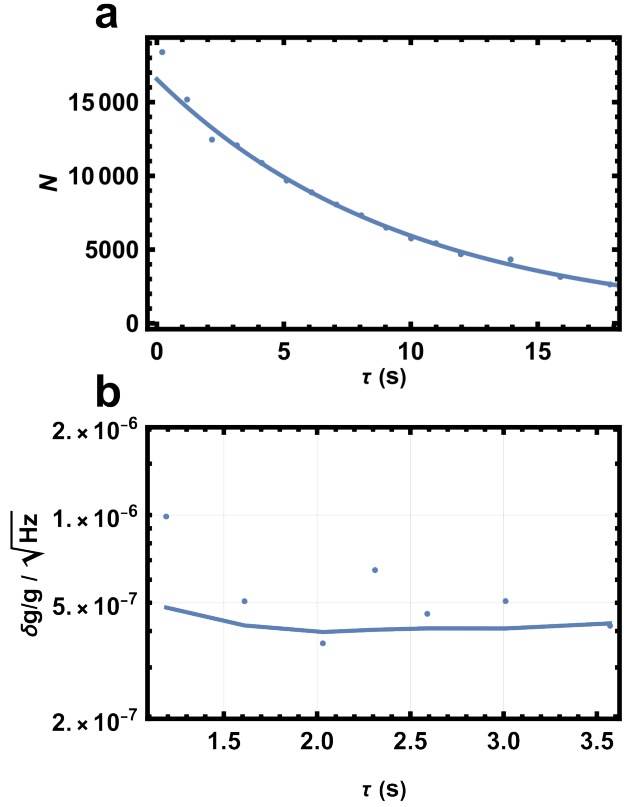

**Extended Data Figure 3. Noise and sensitivity. a,** Measured atom number, $N$, as a function of hold time, $\tau$. This dataset was taken with trap depth $U \approx 12\, E_r$. The solid line shows a fit to a decaying exponential. **b,** Sensitivity and standard quantum limit noise. The measured acceleration sensitivity within one second of measurement time ($\delta g/g/\sqrt{\text{Hz}}$) is shown as dots. Values corresponding to the estimated standard quantum limit (shot) noise are calculated from the measured contrast and atom numbers and shown as solid lines.

Fit parameters and confidence intervals for long $\tau$ fringe contrast

Extended Data Table 1 contains fit parameters for the datasets corresponding to 30s, 60s and 70s that are plotted in Figure 2 in the main text. Column 2 shows the number of shots (images) per dataset. Column 3 shows the average number of detected atoms per shot (see section "Sensitivity and Noise" above for procedure used to extract atom numbers). We note that the atom numbers are several times higher than expected assuming the exponential decay rate shown in Extended Data Figure 3a. This is consistent with the modeling described in the main body of the manuscript, where the rate of atom loss slows down for very long hold times, since the cold atoms that load near the center of the trap are less likely to tunnel than the hot atoms near the edges. The data corresponding to the 70 s fringe was acquired at a later date than the 30 s and 60 s fringes (after atom source upgrades described in the main text), and therefore has a relatively larger atom number per experiment shot and correspondingly better signal to noise.

Columns 4-9 show the resulting fit parameter means $\hat{C}$, $\hat{\phi}$ and $\hat{d}$ and uncertainties $\sigma_{\hat{C}}$, $\sigma_{\hat{\phi}}$ and $\sigma_{\hat{d}}$ from a least-squares non-linear fit to equation $A(\tau) = C \sin(\omega_{\text{prop}} \tau + \phi) + d$. When $C \ll 1$, noise at the standard quantum limits (SQL) limits the contrast error to $\sqrt{2/(N_{\text{total}})}$, where $N_{\text{total}}$ is the total fringe atom number. Corresponding values for each fringe are shown in column 10. The factor of $\sqrt{2}$ comes from the fact that the data is acquired at evenly spaced time intervals, so only half of it is near the extrema of the fringe (in contrast to near the slope of the fringe), where sensitivity to measuring contrast is maximized. We observe excess noise in the experimental data (column 9), consistent with background imaging noise and phase drift, with a smaller, but not insignificant contribution from variations in Earth's gravity due to tides and slow experiment tilts over the hours-long dataset. The correlation coefficients between the three fit parameters, $C_{x-y}$, are shown in columns 12-14 of the table.

**Extended Data Table 1. Long $\tau$ fringe datasets parameters.** Columns show data parameters (#images/datapoint, #atoms/shot), fit parameters ($\hat{\phi}$, $\sigma_{\hat{\phi}}$, $\hat{d}$, $\sigma_{\hat{d}}$, $\hat{C}$, $\sigma_{\hat{C}}$, SQL $\sigma_{\hat{C}}$, and $\hat{C}/\sigma_{\hat{C}}$) and fit correlation coefficients ($C_{C-\phi}$, $C_{C-d}$, $C_{\phi-d}$), while each row is a dataset plotted in Figure 2 corresponding to a particular hold time (30 s, 60 s, 70 s).

| Dataset | #images/datapoint | #atoms/shot | $\hat{\phi}$ | $\sigma_{\hat{\phi}}$ | $\hat{d}$ | $\sigma_{\hat{d}}$ | $\hat{C}$ (%) | $\sigma_{\hat{C}}$ (%) | SQL $\sigma_{\hat{C}}$ (%) | $\hat{C}/\sigma_{\hat{C}}$ | $C_{C-\phi}$ | $C_{C-d}$ | $C_{\phi-d}$ |
|---|---|---|---|---|---|---|---|---|---|---|---|---|---|
| 30 s | 5 | 900 | 1.06 | 0.34 | 0.006 | 0.010 | 4.36 | 1.46 | 0.47 | 3.0 | -0.017 | -0.06 | 0.17 |
| 60 s | 9 | 750 | -0.87 | 0.45 | -0.193 | 0.014 | 4.21 | 1.58 | 0.38 | 2.6 | -0.024 | 0.26 | -0.07 |
| 70 s | 4.5 | 980 | -0.82 | 0.36 | 0.269 | 0.007 | 2.86 | 1.00 | 0.34 | 2.9 | 0.018 | 0.21 | -0.09 |

We use a standard t-test to compute confidence intervals for the hypothesis that the contrast is significantly non-zero for the 30s, 60s and 70s fringes. We compute the t-test statistic by dividing the fitted contrast by the fitted contrast standard error, $\hat{C}/\sigma_{\hat{C}}$, shown in column 11 of Extended Data Table 1. We compute the statistical likelihood that a fringe with no coherence (zero contrast) would result in the measured $\hat{C}/\sigma_{\hat{C}}$. We start by creating 10000 noisy simulated fringes with zero contrast (normally distributed noise consistent with data). We then fit each simulated fringe and create a histogram of the resulting 10000 values of $\hat{C}/\sigma_{\hat{C}}$ (Extended Data Figure 4a). We construct the cumulative distribution function (*cdf*) for $\hat{C}/\sigma_{\hat{C}}$ being nonzero by counting all histogram points where $\hat{C}/\sigma_{\hat{C}}$ is smaller than the *cdf* value (Extended Data Figure 4b).

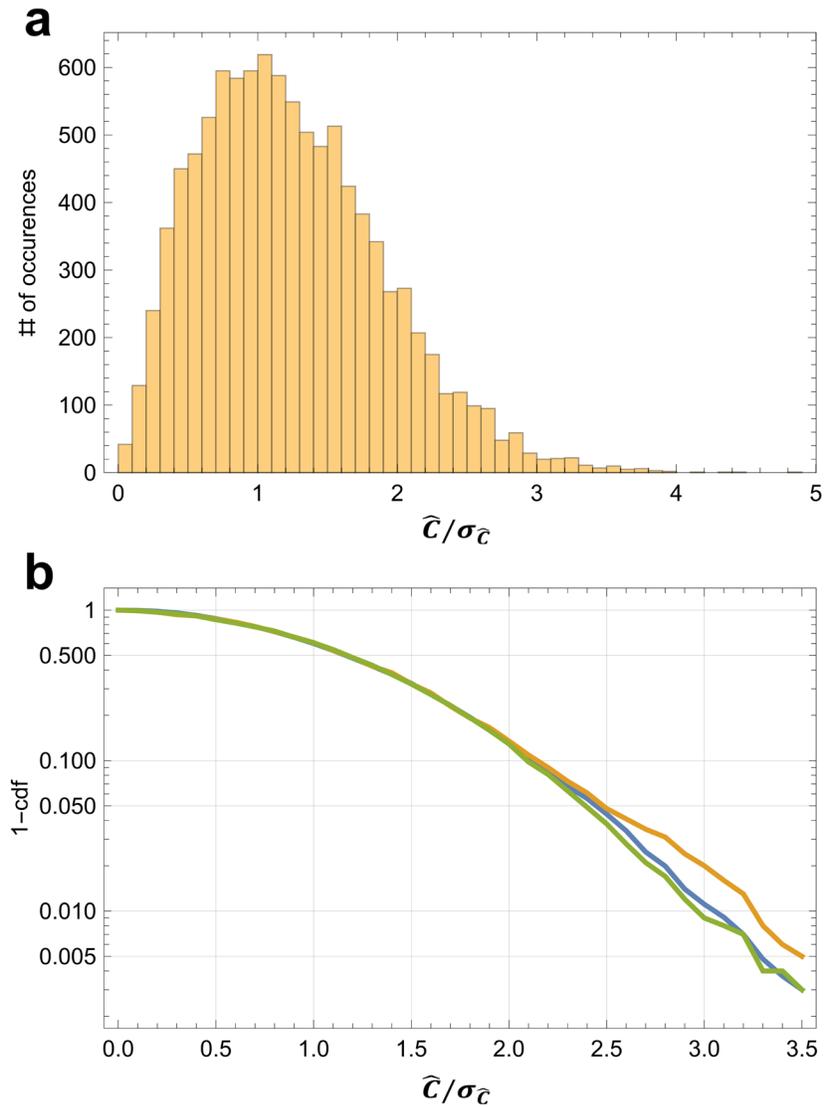

**Extended Data Figure 4. Simulations of the t-test statistic, $\hat{C}/\sigma_{\hat{C}}$. a.** Histogram of the distribution of $\hat{C}/\sigma_{\hat{C}}$ for 1000 simulated fringes with zero contrast $C = 0$. **b.** Cumulative distribution functions (*cdf*) for $\hat{C}/\sigma_{\hat{C}}$, shown for varying levels of simulation added noise (blue –

compatible with experiment imaging and shot noise, yellow-10 times larger, green-10 times smaller).

We use this cumulative distribution function to quote the confidence intervals mentioned in the main text. The ratios of fit mean to fit error shown in Extended Data Table 1 of 3.0, 2.6 and 2.9 for the 30s, 60s and 70s fringes, respectively correspond to confidence of 99%, 97% and 99%, respectively that the contrast extracted from the data is non-zero. Combining the independently acquired 60 and 70 s datasets results in a t-statistic ratio of fit mean to error of 3.9 or 99.9 % likelihood that the fringe data exhibits coherence.

Investigating contrast loss

To find the source of decoherence, we have varied around 20 experimental parameters, looking for a dependence of the contrast decay constant $\kappa$. The choice of many of these parameters is motivated by studies of decoherence in previous experiments with atoms in optical lattices[24–26,38–41]. Most parameters were varied in a range that is large (factor of 2 to 1000) compared to the typical residual value in the experiment. A typical procedure involved measuring experiment fringes both under typical running conditions and when the experimental parameter was greatly exaggerated. Two fringes were taken in each configuration: one that corresponds to low $\Delta z$, $U_0$ and $\tau$, where the contrast approximately equals $C_0$, and one where $\Delta z$, $U_0$ and $\tau$ are set such that $C = C_0/2$, on the slope of the exponentially decaying contrast, $C = C_0 \exp[-\Delta z\, U_0 \tau/\kappa]$. This makes the measurement maximally sensitive to variations in $\kappa$. Taking the ratio of the contrast values of the two fringes gives an accurate estimate of $\kappa$. The list of parameters that were investigated and found to have no influence on contrast decay is given in Extended Data Table 2.

**Extended Data Table 2. Investigating the cause of contrast decay.** Rows list experimental parameters that were varied and found to have no influence on contrast decay at a level of 2%, while columns show the type and outcome of the performed test.

| Parameter Varied | Experimental Test Performed | Observed Outcome |
| --- | --- | --- |
| Radial lattice uniformity | Installed new superpolished cavity mirrors (surface rms <1 A) | Same $\kappa$ |
| Axial lattice uniformity | Replaced planar-concave cavity with symmetric concave-concave cavity (mirrors have equal radii of curvature) | Same $\kappa$ |
| Vacuum pressure | New pumps, reduced outgassing | Same $\kappa$, slower decay of atom number |
| Lattice laser frequency noise | 2x higher with tuning proportional-integral lock | Same $\kappa$ |
| | Narrower laser linewidth by locking to a high-finesse (F=20,000) cavity | |
| Raman beamsplitter symmetry | Symmetric beamsplitters using microwave $\frac{\pi}{2}$ pulse followed by optical $\pi$ pulse *(32)* | Same $\kappa$ |
| Laser lattice broadband emission | Suppressed >10x with filter cavity | Same $\kappa$ |
| Lattice intensity noise | 10x reduced by intensity stabilization using transmitted light as a monitor | Same $\kappa$, slower decay of atom number |
| Background scatter | Increased 20x by shining mode-mismatched light at cavity mirror and/or experiment vacuum window | Same $\kappa$ |
| Acoustic noise | Phone speaker, tapping the optical table | Same $\kappa$, vibrations primarily caused setup translation |
| Alignment with gravity | Tilted optical table by 1.5 mrad | Same $\kappa$, observed expected phase shift due to modified projection of local gravity along cavity axis |
| Changing atom number and density | 2x reduction by lowering state selection efficiency | Same $\kappa$, low contrast at very low atom numbers due to increased imaging noise |
| Axial atom temperature selection | Reduced 3x by increasing the length of velocity selection pulse | Same $\kappa$, as expected since we only expect dependence on radial temperature |
| Misaligned lattice laser, coupling light to high order cavity modes | Reduced cavity coupling efficiency 2x by misaligning and changing beam diameter | Same $\kappa$ |
| Lattice laser detuning | Replaced lattice laser with 866 nm ECDL (14 nm det.) and 1064 nm fiber laser (212 nm det.) | Same $\kappa$, except change in contrast due to single photon scattering, as expected |
| Environmental field gradients | Varied the vertical atom position by up to 1.5 cm | Same $\kappa$ |

| Magnetic field gradients | Increased 1000x by turning on MOT coils during interferometer | Same $\kappa$, observed expected phase shift due to quadratic Zeeman shift |
|---|---|---|
| Position within the atomic sample | Analyzed horizontal and vertical slices of fluorescence image | Same $\kappa$ |

Loading atoms into higher order modes

Supplementary Movie 1 displays a sequence of fluorescence images showing the atom cloud being loaded into the horizontally asymmetric Hermite-Gauss TEM01 mode of the optical cavity.

To produce data shown in Fig. 3c of the main text, we use higher order cylindrically symmetric Laguerre-Gauss modes (LG10, LG20), which have increasingly narrower mode diameters of the central peak. Lattice laser light is coupled into these higher order modes by increasing the diameter of the incoming laser beam such that the incoming Gaussian beam has good overlap with the LG modes (typically 20-30%). An acousto-optic modulator (AOM) is used to sweep the lattice laser frequency (sweep time 1 ms) over the optical cavity transverse mode spacing, switching between coupling the lattice laser to the LG00 and either LG10 (20 MHz sweep) or LG20 (40 MHz sweep range).

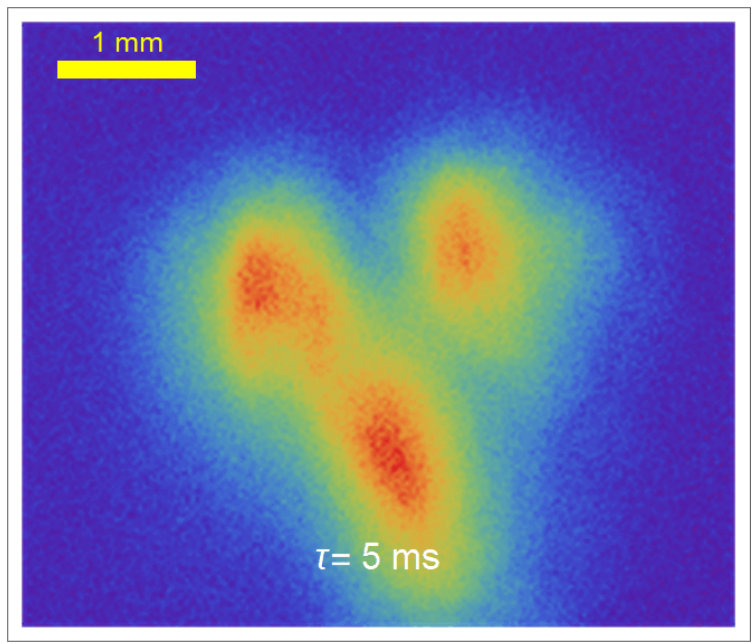

**Supplementary Movie 1. Sequence of fluorescence images showing dynamics of the spherically symmetric atom cloud loading into a TEM01 mode of the optical cavity.** The yellow and red colors represent regions of space with high atom density. The atoms that have good overlap with the cavity get loaded and initially oscillate (period 150 ms) before eventually (after 300-400 ms) filling the available cavity mode trap volume.

Lattice trajectories and contributions to lattice phase shift

The diagram shown in Extended Data Figure 6 is a pictorial representation of the atomic trajectories inside the optical lattice and resulting differential phase shifts between the top and bottom interferometer wavepackets due to oscillatory tilts. It complements the description of decoherence due to such effects in the main text.

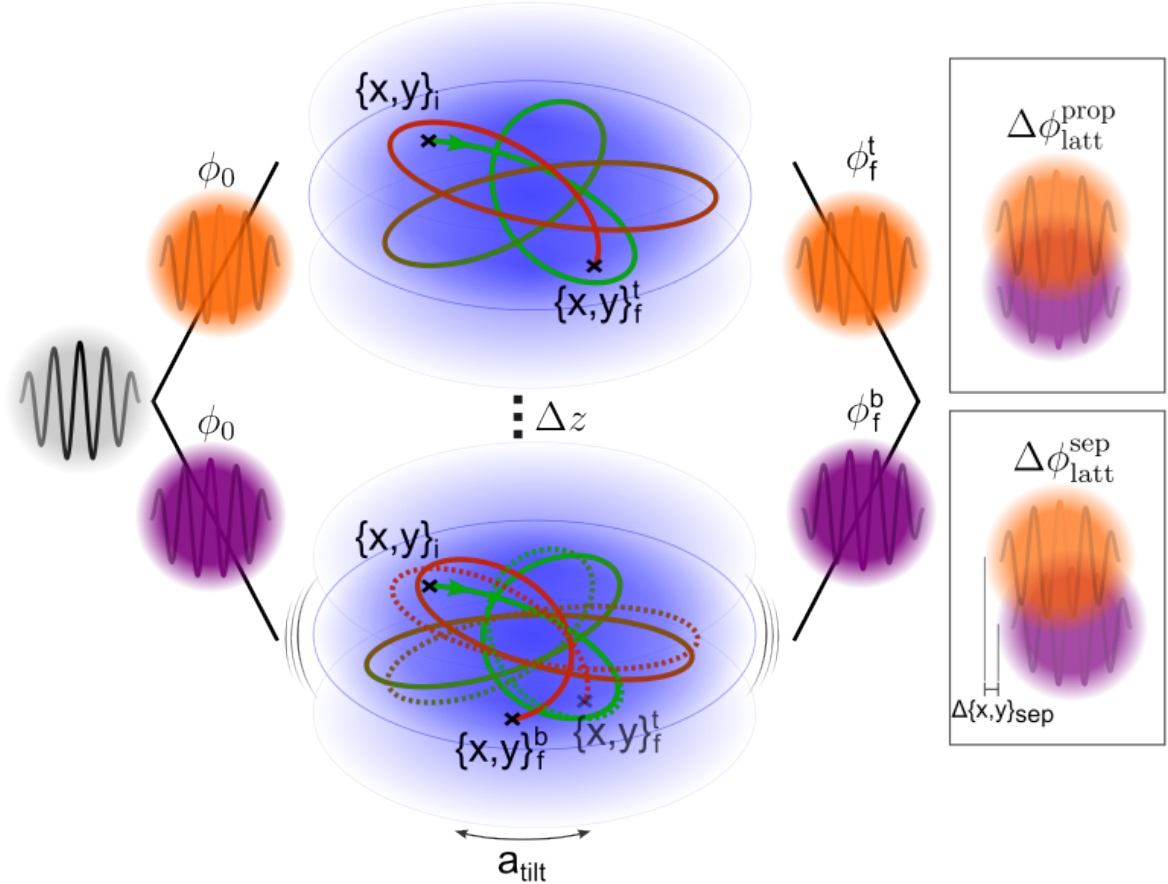

**Extended Data Figure 6. Phase shifts due to differential motion between the atomic wavepacket in the top and bottom lattice sites.** The atomic wavepacket (wavy line on top of grey disk) is split into an equal superposition of two partial wavepackets (wavy lines on top of purple and orange disks). The partial wavepackets are loaded into lattice sites (blue ovals) that are spatially separated by a distance of $\Delta z$. Solid lines show representative atom trajectories in the two lattice sites, which are both initially loaded at position $\{x, y\}_i$. Survival probabilities due to Landau-Zener tunneling are shown by the trajectories color gradients and range from 1 (green) to 0 (red). Time-dependent tilts cause the position of the lower lattice site to oscillate with a differential amplitude relative to the top in the transverse plane, $\Delta a_{\text{tilt}}$, leading to a displaced trajectory of the bottom atom partial wavepacket with final position $\{x, y\}_f^t \neq \{x, y\}_f^b$ and propagation phase difference $\phi_f^t \neq \phi_f^b$. The rightmost panels show the two main dephasing mechanisms, due to differential propagation phase, $\Delta\phi_{\text{latt}}^{\text{prop}}$, and separation phase, $\Delta\phi_{\text{latt}}^{\text{sep}}$.

Numerical simulation

As mentioned in the main text, we use a simulation with parameters that match our knowledge of the experiment (Extended Data Table 3) to estimate the effect of lattice imperfections on interferometer decoherence. The simulation starts with a sample of 500-1000 atoms generated with 2D positions and velocities (in the $xy$ plane) randomly drawn from Gaussian distributions with diameters of 1520 $\mu$m and 13 mm/s, respectively.

Each atom's trajectory parameters $\{x(t), y(t), v_x(t), v_y(t)\}$ are calculated by integrating the equations of motion in each $x, y$ spatial direction

$$\begin{aligned}\frac{dx(t)}{dt} &= v_x(t)\\ \frac{dy(t)}{dt} &= v_y(t)\\ \frac{dv_x(t)}{dt} &= -\frac{dU_{\text{latt}}(x,y,t)}{dx}\frac{1}{m_{\text{Cs}}}\\ \frac{dv_y(t)}{dt} &= -\frac{dU_{\text{latt}}(x,y,t)}{dy}\frac{1}{m_{\text{Cs}}},\end{aligned} \quad (6)$$

where $U_{\text{latt}}(x,y,t) = 0$ for the first 38 ms, corresponding to free-propagation and then $U_{\text{latt}}(x,y,t) = U_0 \text{Exp}\left(-\frac{(x-x_0(t))^2 + (y-y_0(t))^2}{2w^2}\right)$ during the lattice hold, where the center of the Gaussian potential is set to $x_0(t) = y_0(t) = 0$ for the unperturbed potential. In addition to trajectory parameters, probabilities for the atoms to leave the trap $p(t)$ are calculated from the Landau-Zener criterion[42] through $\frac{dp}{dt} = \ln\{1 - \text{Exp}\left[-\frac{\pi}{64}\frac{a_{\text{Cs}}}{g}\left(\frac{U_{\text{latt}}(x(t),y(t),t)}{E_r}\right)^2\right]\}$, where $a_{\text{Cs}} = \hbar^2 k^3 / m_{\text{Cs}}^2$. An atom that tunneled is assumed to have left the experiment and does not contribute to the interferometer output. The interferometer propagation phase is calculated from $\phi_{\text{latt}}^{\text{prop}}(t) = 1/\hbar \int \mathcal{L}\, dt$, where $\mathcal{L} = \frac{1}{2}m_{\text{Cs}}(v_x(t)^2 + v_y(t)^2) - U_{\text{latt}}(x(t),y(t),t)$. The integration uses a Runge-Kutta routine, where the timestep (typically 100 $\mu$s) is much shorter than the timescales of the motion (10s of Hz). We verify that simulation results are independent of timestep size choice within an order of magnitude.

**Extended Data Table 3. Simulation parameters.** Rows list simulation parameters, while columns show the typical parameter numerical values used.

| Simulation Parameter | Value |
| --- | --- |
| Atom sample x, y, z position spread diameter | 1520 $\mu$m |
| Atom sample x, y velocity spread diameter | 13 mm/s |
| Free-propagation time before hold | 38 ms |
| Tilt vibration frequency $\omega_{\text{tilt}}$ | $2\pi \cdot 200$ Hz |
| Tilt vibration amplitude $\theta$ | $0.3 * 10^{-3}$ rad |
| Vibration pivot arm $L$ | 0.75 m |
| Integration time step size | 100 $\mu$s |

The trajectory positions and velocities, $p(t)$ and $\phi_{\text{latt}}^{\text{prop}}(t)$ are each calculated independently for the top and bottom interferometer arms. The effect of tilts with amplitude $\theta$ is modeled by moving the center of the lattice potential to a time-dependent position $x_0^{\{t,b\}}(t) =$ $a_{\text{tilt}}^{\{t,b\}} \sin(\omega_{\text{tilt}} t)$, where $a_{\text{tilt}}^{t} = \theta (L + \Delta z)$ is the top arm potential oscillation amplitude, while the bottom arm oscillation amplitude is $a_{\text{tilt}}^{b} = \theta L$. The vibrations lever arm $L = 75$ cm and amplitude $\theta = 0.3 \cdot 10^{-3}$ radians are consistent with vibrations of the experimental frame. The oscillation frequency is set to $\omega_{\text{tilt}} = 2\pi \cdot 200$ Hz, which is arbitrary since we find that the decoherence rate is independent of $\omega_{\text{tilt}}$ as long as $\omega_{\text{tilt}}$ is larger than the trap frequency (~few Hz).

At the end of the integration, the separation phase is calculated from the final atom positions and velocities as $\phi_{\text{latt}}^{\text{sep}}(t) = [(x^t - x^b)(v_x^t + v_x^b) + (y^t - y^b)(v_y^t + v_y^b)]m_{\text{Cs}}/(2\hbar)$, where superscripts $t$ and $b$ correspond to positions and velocities of the top and bottom interferometer arms. The total differential phase between the top and bottom arms is then $\Delta\phi_{\text{latt}}(t) = \Delta\phi_{\text{latt}}^{\text{prop}}(t) + \phi_{\text{latt}}^{\text{sep}}(t)$, where $\Delta\phi_{\text{latt}}^{\text{prop}}(t)$ is the difference in propagation phase between the top and bottom arms.

The average ensemble phase shift $\langle\Delta\phi_{\text{latt}}\rangle$ and phase spread, $\delta(\Delta\phi_{\text{latt}})$ are calculated by performing a sum over all atomic trajectories. The effect of the phase spread $\delta(\Delta\phi_{\text{latt}})$ on the fringe contrast $C$ is quantified by computing an average of fringes with phases given by the ensemble phase distribution, each weighted by their individual Landau-Zener survival probabilities, $p(t)$. A third contribution to fringe contrast loss due to poor overlap of the atomic wavepackets at the end of the interferometer is found to be negligible compared to the dephasing described above. The contrast decay time constant $\tau_C$ is extracted from a decaying exponential fit (where $\tau_C$ is the only free parameter) of the fringe contrast sampled for $t = 0$ to $t = 20$ s in steps of 0.2 s. The simulation time was extended to 70 s for generating Figure 3a. We find good agreement between the simulations and experimental data (Figure 3).

We verify the performance of the trajectory simulation by measuring the lattice loading fraction under different laboratory conditions and comparing the resulting atom numbers with the simulation, as a fraction of the initially available atoms. We find good agreement between the atom numbers loaded and remaining in the lattice after a 1 second hold, when the experiment and simulation tilt is varied (Extended Data Figure 5a) and when the trap depth if varied (Extended Data Figure 5b).

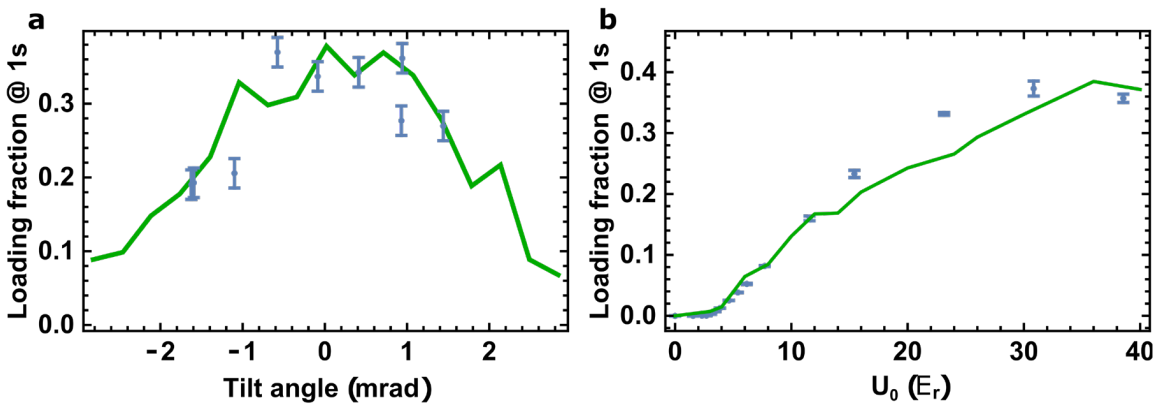

**Extended Data Figure 5. Loading fractions.** Experimental data showing fractions of atoms remaining after loading and 1 second of hold time as a function of initial number, shown vs **a,** global tilt angle and **b,** trap depth, $U_0$. Experimental data is shown by blue dots with error bars corresponding to $1\sigma$ (68% confidence) Gaussian intervals, and simulation results are in green.

We also investigate contrast loss as a function of hold time $\tau$ for various simulation configurations (see Extended Data Figure 7): with the standard configuration described above, with tilt vibrations set to zero, with Landau-Zener tunneling disabled, and with 5-fold reduced velocity and position distributions (lower temperature). In contrast to the slowing of the contrast decay observed for $\tau > 20$ s with standard simulation conditions (blue band, Extended Data Figure 7), we observe contrast loss that continues along an exponential trend when Landau-Zener tunneling is disabled (green band, Extended Data Figure 7). This is consistent with the model put forward in the main text, as explained in the following. Without tunneling, the atom sample distribution is no longer energy selected by the optical lattice and high energy atoms remain in the sample. These atoms continue to sample the large potential gradients present near the edges of the trap and are therefore more likely to de-phase and contribute to relatively strong exponential contrast loss. We also run simulations with an atom sample which has 5-fold narrower velocity and position distributions (red band, Extended Data Figure 7), which further confirms that atoms with lower energy produce greatly reduced decoherence.

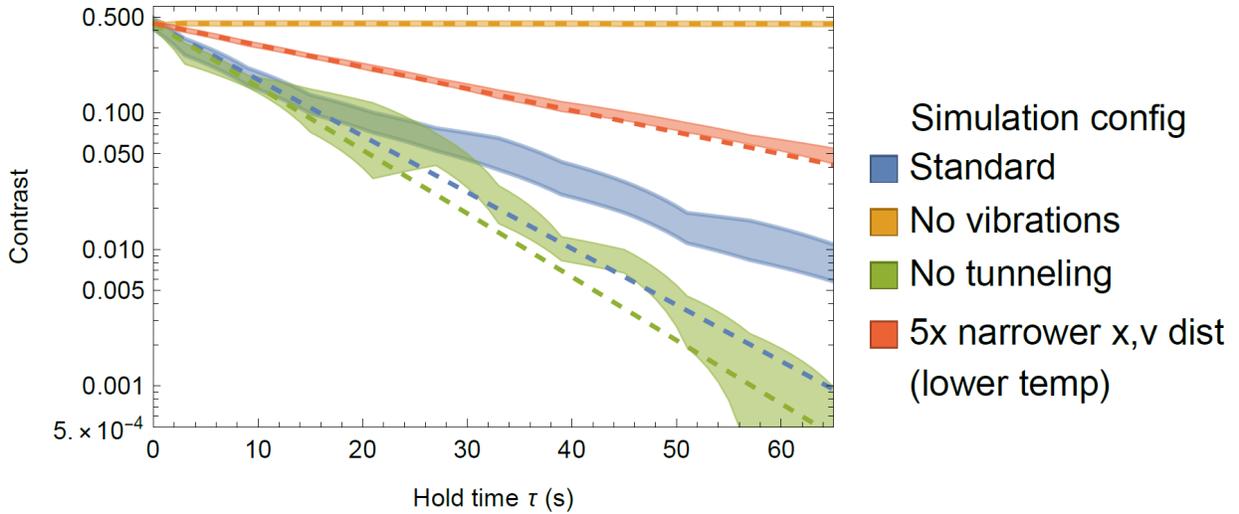

**Extended Data Figure 7. Simulated contrast decay.** Contrast vs hold time $\tau$ for four separate configurations: standard level of vibrations, with vibrations reduced to zero, with tunneling reduced to zero, and with atom position and velocity distributions widths reduced 5-fold. All simulations used a wavepacket separation of $\Delta z = 1.9\ \mu m$ and a peak trap depth of $U_0 = 7\ E_r$. The dashed lines represent exponential fits to the $\tau < 20$ s region of the data. The bands represent 95 % confidence intervals.

Density matrix formalism

We use a density matrix formalism[33] to show that a Lorentzian distribution in atom phases, $\delta(\Delta\phi_\text{latt})$, results in the exponential decay of the ensemble contrast observed experimentally, a result that is used in the main text Eq. 5. Initially, the atoms are in a coherent superposition state of the top and bottom arms: $|\psi\rangle_i = (|t\rangle + |b\rangle)/\sqrt{2}$. This corresponds to a pure state density matrix:

$$\rho_i = |\psi\rangle_i\langle\psi|_i = \frac{1}{2}\begin{pmatrix} 1 & 1 \\ 1 & 1 \end{pmatrix}. \tag{7}$$

While in the optical lattice, the lower lattice site accumulates phase $\Delta\phi_{\text{latt}}$ and the state evolves as: $|\psi\rangle_f = (|t\rangle + e^{i\Delta\phi_{\text{latt}}}|b\rangle)/\sqrt{2}$, corresponding to evolution operator $\mathcal{E}(\Delta\phi_{\text{latt}}) = \begin{pmatrix} 1 & 0 \\ 0 & e^{i\Delta\phi_{\text{latt}}} \end{pmatrix}$. The final density matrix of the two-level system is given by $\rho_f = \int_{-\infty}^{\infty} \mathcal{E}(\phi)\rho_i\mathcal{E}^{\dagger}(\phi)P(\phi)\,d\phi$, which evaluates to

$$\rho_f = \frac{1}{2}\begin{pmatrix} 1 & \text{Exp}[-\delta(\Delta\phi_{\text{latt}})] \\ \text{Exp}[-\delta(\Delta\phi_{\text{latt}})] & 1 \end{pmatrix}. \tag{8}$$

This represents a mixed state where the off-diagonal terms describe the coherence of the system. The contrast thus decays exponentially with increasing phase spread

$$C = C_0 \text{Exp}[-\delta(\Delta\phi_{\text{latt}})]. \tag{9}$$

**Data availability:** All data presented in this paper are deposited online[43].

**Code availability:** All code used in this paper is available upon reasonable request.

### Methods-only references: